\RequirePackage{lineno}
\documentclass[prd, onecolumn, amsmath,amssymb,superscriptaddress,footinbib,12pt]{revtex4}
\usepackage{url}
\usepackage{lineno}

\usepackage{epsfig}
\usepackage{amsfonts}
\usepackage{graphicx}
\usepackage{epsfig}
\usepackage{eepic}
\usepackage{amsmath}
\usepackage{amssymb}
\usepackage{color}
\usepackage{bbm}
\usepackage{dcolumn}
\usepackage{bm}
\usepackage{ulem}
\usepackage{mathrsfs}
\usepackage{bbold}
\usepackage{datetime}

\usepackage{overpic} 
\usepackage{rotating}
\usepackage[usenames,dvipsnames]{xcolor}
\usepackage[colorlinks=true,citecolor=Magenta,linkcolor=Green,urlcolor=Green]{hyperref}
\usepackage{lipsum} 
\usepackage{tikz,tikz-3dplot}
\usetikzlibrary{shapes.geometric}
\usepackage{etoolbox} 
\usepackage[capitalize]{cleveref}
\usepackage{extarrows}

\def\bc{\begin{center}}

\def\ec{\end{center}}
\def\be{\begin{eqnarray}}
\def\ee{\end{eqnarray}}
\definecolor{dyellow}{rgb}{1.,0.8,.0}
\definecolor{myblue}{rgb}{.1,.1,.7}
\definecolor{dcyan}{rgb}{.0,.6,.6}
\definecolor{dmagenta}{rgb}{0.6,0.0,0.6}
\definecolor{brown}{rgb}{0.6,0.2,0.}
\definecolor{darkblue}{rgb}{.0,.0,0.5}
\definecolor{darkred}{rgb}{0.75,0.0,0.0}
\definecolor{orange}{rgb}{1.,.6,.0}
\definecolor{dorange}{rgb}{0.8,.4,.0}
\definecolor{darkgreen}{rgb}{0.0,0.6,0.0}
\definecolor{purple}{rgb}{.4,.0,.4}
\definecolor{lightgrey}{rgb}{0.7, 0.7, 0.7}
\definecolor{grey}{rgb}{0.4, 0.4, 0.4}

\usepackage[position=t, singlelinecheck=off]{subfig}
\usepackage[font=small,labelfont=bf,justification=raggedright]{caption}

\newcommand{\xdownarrow}[1]{%
  {\left\downarrow\vbox to #1{}\right.\kern-\nulldelimiterspace}
}
\newcommand{\xuparrow}[1]{%
  {\left\uparrow\vbox to #1{}\right.\kern-\nulldelimiterspace}
}

\definecolor{myred}{RGB}{189, 38, 49}

\begin{document}

\title{Periodicities in a multiply-connected geometry from quenched dynamics}
\author{Zhi-Hong \surname{Li}} \email{lizhihong@buaa.edu.cn}
\affiliation{Center for Gravitational Physics, Department of Space Science, Beihang University,
Beijing 100191, China}
\author{Hai-Qing Zhang} \email{hqzhang@buaa.edu.cn}
\affiliation{Center for Gravitational Physics, Department of Space Science, Beihang University,
Beijing 100191, China}
\affiliation{International Research Institute for Multidisciplinary Science, Beihang University, Beijing 100191, China}

\begin{abstract}
{Exploring the lowest energy configurations of a quantum system is consistent with the counting statistics of the frequently appeared states from quenching dynamics. By studying the Little-Parks periodicities in a multiply-connected ring-shaped geometry from the holographic technique, it is found that the frequently appeared states from dynamics incline to have lower free energies.  In particular, the resulting winding numbers from quenched dynamics are constrained in a normal distribution for a fixed magnetic flux threading the ring. Varying the magnetic fluxes, Little-Parks periodicities will take place with periods identical to the flux quantum $\Phi_0$. Favorable solutions with lowest free energies perform first order phase transitions which transform between distinct winding numbers as the magnetic flux equals half-integers multiplying $\Phi_0$.  }

\end{abstract}

\maketitle


\section{Introduction}
It is a fundamental problem in quantum systems to probe the relations between the frequently appeared states from dynamics and the states with lowest free energies in equilibrium. This problem bridges the gaps between the dynamical properties, statistical mechanics and equilibrium physics in a quantum system \cite{Bogolubov}. For this purpose, we conceive a model of Little-Parks (LP) periodicities by quenching a disordered system into a symmetry-breaking state by virtue of the technique of gauge-gravity duality \cite{Maldacena:1997re}, and find that the frequently appeared states from dynamics incline to have lower free energies in equilibrium. 

LP experiment \cite{LP} is a hallmark of demonstrating the pairing of electrons in the Bardeen-Cooper-Schrieffer (BCS) superconductors \cite{BCS}.  The experiment is composed of a hollow thin-walled superconducting cylinder threading by an axial magnetic field. The critical temperature turns out to be a periodic function of the flux quantum $\Phi_0=2\pi\hbar c/e^*$ enclosed by the cylinder, where $e^*$ is the charge of the electron pairs \cite{tinkham}. From Ginzburg-Landau theory, this periodicity arises from the existence of a single-valued complex wave function in the multiply connected geometry \cite{tinkham}. LP effect has been widely extended to, for instance, an inhomogenous superconducting cylinder \cite{nikulov}, a small ring geometry \cite{Vakaryuk,Loder}, a two-band superconductor \cite{erin}, and etc.  Extensions of LP effect to strongly coupled systems were carried out by virtue of the gauge-gravity duality in \cite{Montull:2011im,Montull:2012fy,Cai:2012vk}, however, these work was done in static and the winding numbers were brought in by hand. 

We will generate the LP periodicities dynamically, in particular the winding numbers of the order parameter will turn out automatically and stochastically at the final equilibrium state. A natural way to simulate the dynamical process resorts to the Kibble-Zurek mechanism (KZM) \cite{Kibble:1976sj,Kibble:1980mv,Zurek:1985qw}, in which quenching a system from higher symmetries across the critical point, topological defects will arise in the symmetry-breaking phase. KZM has been widely tested in numerous systems \cite{Chuang:1991zz,Ruutu:1995qz,Carmi:2000zz} and was supported by various numerical studies \cite{Laguna:1996pv,Yates:1998kx,Donaire:2004gp}.  In particular, holographic studies of KZM in spatial 1D and 2D systems were carried out in \cite{Sonner:2014tca,Chesler:2014gya,Zeng:2019yhi,Li:2019oyz,delCampo:2021rak,Li:2021iph,Li:2021dwp,Xia:2021xap,Li:2021jqk}. Reviews can be referred to \cite{kibblereview,zurekreview}. 

In this paper, we study the dynamical superconducting transition in a spatially 1D ring by decreasing the temperature across the critical point. The topological defects -- winding numbers -- will turn out due to the KZM \cite{Das:2011cx,Xia:2020cjl}.  The winding number is a topologically invariant quantity due to the single valuedness of the order parameter around the ring.  For a fixed value of the threading magnetic flux $\Phi$, the statistics of winding numbers are constrained in a normal distribution with the mean identical to the ratio $\Phi/\Phi_0$. By varying the magnetic fluxes, the conserved currents, average absolute values of the order parameter and the free energies all display periodic relations with respect to the magnetic fluxes, with the identical periods equalling $\Phi/\Phi_0=1$. These are typical results of LP effect in a thin-walled cylinder. The favorable solutions with lower free energies display a first order phase transition at the half-integers of $\Phi/\Phi_0$. By comparing the analysis of the free energies and the distributions of the winding numbers, we demonstrate that the frequently appeared states from dynamics incline to have lower free energies, which are thermodynamically favorable. This work opens up a possibility of exploring the lowest energy configurations of a quantum system from the quenched dynamics and the statistical distributions of the final states.

\section{Holographic Mapping}
The holographic superconducting system can be simulated by the Abelian-Higgs model in the bulk \cite{Hartnoll:2008vx}, 
\begin{eqnarray}\label{density}
\mathcal{L} = -\frac{1}{4} F_{\mu \nu} F^{\mu \nu} - |D \Psi|^2 - m^2 |\Psi|^2,
\end{eqnarray}
where $F_{\mu\nu}=\partial_\mu A_\nu-\partial_\nu A_\mu$ is the field strength of the U(1) gauge field $A_\mu$,  $\Psi$ is the complex scalar field and $D_\mu=\nabla_\mu -iA_\mu$ is the covariant derivative. (The units $\hbar=c=e^*=1$ were adopted.) We work with the Eddington-Finkelstein coordinates in AdS$_4$ planar black hole to inspect the temporal dynamics of the holographic superconducting system \cite{Chesler:2013lia},
\begin{eqnarray}
ds^2 = \frac{1}{z^2} \left(-f(z) dt^2 - 2dtdz + dx^2 +dy^2 \right),
\end{eqnarray}
where $f(z) = 1 - (z/z_h)^3$ and $z_h$ represents the horizon location. (We have set the AdS radius $l=1$.) The AdS boundary is at $z = 0$ and the horizon can be scaled as $z_h=1$. Therefore, the Hawking temperature is given by $T=3/(4\pi)$. 
In the probe limit, the equations of motions read,
\begin{eqnarray}\label{eomofwhole}
D_\mu D^\mu\Psi-m^2\Psi=0, \qquad \nabla_\mu F^{\mu\nu}=i\left(\Psi^* D^\nu\Psi-\Psi{(D^\nu\Psi)^*}\right). 
\end{eqnarray}
Holographically, we propose a toy model in the boundary field theory to mimic the LP experiment, which consists of a thin-walled cylinder threading by an axial magnetic field. That is, we set the periodic boundary conditions of all the fields along $x$-direction to represent the compact ring, and impose the homogeneous dependence of fields on the $y$-direction indicating the axial direction of the cylinder. Therefore, the ansatz for the fields are $\Psi = \Psi(t,z,x), A_{t} = A_{t}(t,z,x), A_x=A_x(t,z,x)$ and $A_z =A_y= 0$. 
And the explicit forms of the equations read, 
 \begin{eqnarray}
\label{eompsi}
\partial_t \partial_z \psi  - \frac12 \big[( i \partial_z A_t -z-i \partial_x A_x - A_x^2 )\psi  + (f' +2 i A_t )\partial_z \psi  + f \partial_z^2 \psi - 2 i A_x \partial_x \psi+ \partial_x^2 \psi  \big] = 0;~&
\\
\label{eom2}
\partial_t \partial_z A_t - \partial_x( \partial_x A_t  + f \partial_z  A_x  - \partial_t  A_x) 
+ 2 A_t |\psi|^2 + 2\Im\left(\psi\partial_t\psi^*-f\psi\partial_z\psi^*\right)  = 0;~&
\\
\label{eom3}
\partial_t \partial_z A_x - \frac12 \big[ \partial_z (\partial_x A_t + f \partial_z A_x) -2\Im\left(\psi\partial_x\psi^*\right) \big] +  A_x |\psi|^2 = 0;~&
\\
\label{eom1}
\partial_z^2A_t-\partial_z \partial_x A_x  -2\Im\left(\psi\partial_z\psi^*\right) = 0.~&
\end{eqnarray}
where $\psi=\Psi/z$, $f'=\partial_zf$ and the symbol $\Im(\dots)$ indicates the imaginary part.
The above four equations are not independent, indeed the left hand side of the above equations satisfy the following constraint equation,
\begin{eqnarray}
\frac{d}{dt}\text{Eq.\eqref{eom1}}+\frac{d}{dz}\text{Eq.\eqref{eom2}}-2\frac{d}{dx}\text{Eq.\eqref{eom3}}\equiv 2i\left(\text{Eq.\eqref{eompsi}}\times\psi^*-\text{Eq.\eqref{eompsi}}^*\times\psi\right).
\end{eqnarray}
Therefore, there are three independent equations for three fields,  $\psi, A_t$ and $A_x$. Since $\psi=\Psi/z$ is a complex field, this also means that there are four independent real fields for four independent real equations. It in turn implies that our choice of the gauge $A_z=A_y=0$ is viable for the setup of the system.

Without loss of generality, we set the scalar mass square as $m^2= -2$ \footnote{In the framework of holography, the value of the mass square $m^2$ is related to the conformal dimension of the scalar operator in the boundary field theory. We set $m^2=-2$ since in this case the expansion of the scalar field $\Psi$ near the the boundary is like $\Psi|_{z\to0}\sim \Psi_0 z +\Psi_1z^2$. The powers of $z$ in this expansion make it easier to be computed numerically, see for instance in Ref.\cite{Hartnoll:2008vx}. Of course,  different mass squares (or different conformal dimensions $\Delta$) will surely affect some properties, such as the values of the condensate of the order parameter in the holographic superconductor. However, it does not change the general properties of formations of the winding number, i.e.  the Kibble-Zurek mechanism still holds, as we see in Ref.\cite{Sonner:2014tca}, where the authors chose $m^2 = 0$ to study the formation of winding numbers in a 1D ring holographically.}. The asymptotic behaviors of fields near $z\to0$ are 
\begin{eqnarray}
A_\mu&\sim& a_\mu+b_\mu z+\dots,\\
\Psi&\sim& \Psi_0 z^{\Delta_-}+\Psi_1 z^{\Delta_+}+\dots, \ \ \ 
{\rm where}\ \ \Delta_{\pm}=\frac{1}{2}(3\pm\sqrt{9+4m^2}).
\end{eqnarray}
in which $\Delta_{\pm}$ is the conformal dimension of the dual scalar operator in the boundary field theory. As $m^2= -2$, $\Delta_-=1$ and $\Delta_+=2$. Then the asymptotic behavior of scalar field near $z\to0$ is $\Psi= z\left(\Psi_0+\Psi_1 z+\dots\right)$, which is easy to be calculated numerically. 
From the dictionary of gauge-gravity duality, $a_t, a_x$ and $\Psi_0$ are interpreted as the chemical potential, potentials of the spatial component of gauge fields and source of scalar operators on the boundary, respectively. Correspondingly, $b_t, b_x$ and $\Psi_1$ are related to the charge density $\rho$, the conserved current $J_x$ and condensate of the order parameter $\langle O\rangle$. At the horizon $z_h$, we set $A_t(z_h)=0$ and the regular finite boundary conditions for other fields. At the boundary $z\to0$, we impose $\Psi_0=0$ in the standard quantization \cite{Hartnoll:2008vx} and set the Dirichlet boundary conditions for $A_x$.

\section*{\bf Methods}\label{methods}
{\bf Holographic renormalization \& holographic free energy -} From holographic renormalization \cite{Skenderis:2002wp}, the holographic free energy can be computed from the on-shell action and the corresponding counter-terms. The generic on-shell action of the Lagrangian \eqref{density} is \cite{Garcia-Garcia:2013rha}
\begin{eqnarray}
S_{\text{os}}&=&-\frac12\int d^4x\partial_\mu\left[\sqrt{-g}\left(A_\nu F^{\mu\nu}+\Psi^*\partial^\mu\Psi+\Psi\partial^\mu\Psi^*\right)\right]\nonumber\\
&&+\frac{i}{2}\int d^4x\sqrt{-g}A_\mu\left(\Psi^*D^\mu\Psi-\Psi(D^\mu\Psi)^*\right). 
\end{eqnarray}
The counter-term in this case is $S_{\rm ct}=\int d^3x\sqrt{-h}\Psi^*\Psi$, where $h$ is the reduced metric on the $z\to0$ boundary. The finite renormalized on-shell action $S_{\rm re}$ can be obtained by $S_{\rm re}=S_{\text{os}}+S_{\rm ct}$.  The free energy is $F=-TS_{\rm re}$ where $T$ is the temperature of the black hole. Replacing the fields and the corresponding boundary conditions into the free energy, we finally arrive at
\be
F&&=-\frac12 V_y\int dx\left(A_t\partial_zA_t-A_x\partial_z A_x\right)\bigg|_{z=0}\nonumber\\
&&-V_y\int dzdx\left(A_t\text{Im}\left(\psi\partial_z\psi^*\right)-A_x\text{Im}\left(\psi\partial_x\psi^*\right)-A_x^2|\psi|^2\right).~~~~
\ee
where $V_y=\int dy$ is the volume along $y$-direction and $\psi=\Psi/z$.

{\bf Numerical schemes -} From the dimensional analysis, the temperature of the black hole $T$ has mass dimension one, while the mass dimension of the charge density $\rho$ on the boundary is two. Therefore, $T/\sqrt{\rho}$ is a dimensionless quantity. From the holographic superconductor \cite{Hartnoll:2008vx}, reducing the temperature is equivalent to increasing the charge density. Thus, in order to linearly quench the temperature as $T(t)/T_c=1-t/\tau_Q$ from KZM \cite{Kibble:1976sj,Kibble:1980mv,Zurek:1985qw} ($\tau_Q$ is the so called quench rate), one can indeed quench the charge density $\rho$ as
$\rho(t)={\rho_c}/{\left(1-t/\tau_Q\right)^{2}}$,
where $\rho_c$ is the critical charge density in the homogeneous and static holographic superconducting system \footnote{In the
 Appendix \ref{app}, we use the Sturm-Liouville eigenvalue problems to analytically get the periodic relations between the critical phase transition points $\rho_c$ with respect to $\Phi/\Phi_0$. }.
 Before quenching, we thermalize the system thoroughly in order to make an equilibrium initial state. To this end, we add the Gaussian white noise $\xi(x_i,t)$ to each field, i.e, $A_t, A_x$ and $\psi$, in the bulk with $\langle \xi(x_i,t)\rangle=0$ and $\langle \xi(x_i,t)\xi(x_j,t')\rangle=h\delta(t-t')\delta(x_i-x_j)$, where we choose $h$ as a small amplitude $h=10^{-3}$. Afterwards, we linearly quench the temperature from $T_i = 1.4T_c$ to $T_f = 0.8T_c$, the system evolves from a normal metal state to a superconductor state.  We evolve the system by using the 4th-order Runge-Kutta method with time step $\Delta t=0.1$. In the AdS radial direction $z$, we adopt the Chebyshev  pseudo-spectral methods with 21 grid points. Since all the fields are periodic in the $x$-directions, we use the Fourier decomposition along $x$-directions with $201$ grid points.\\

\section{Results}
\subsection{Quenched dynamics}
\begin{figure}[th]
\includegraphics[trim=0cm 0cm 1.2cm 0.cm, clip=true, scale=0.31, angle=0]{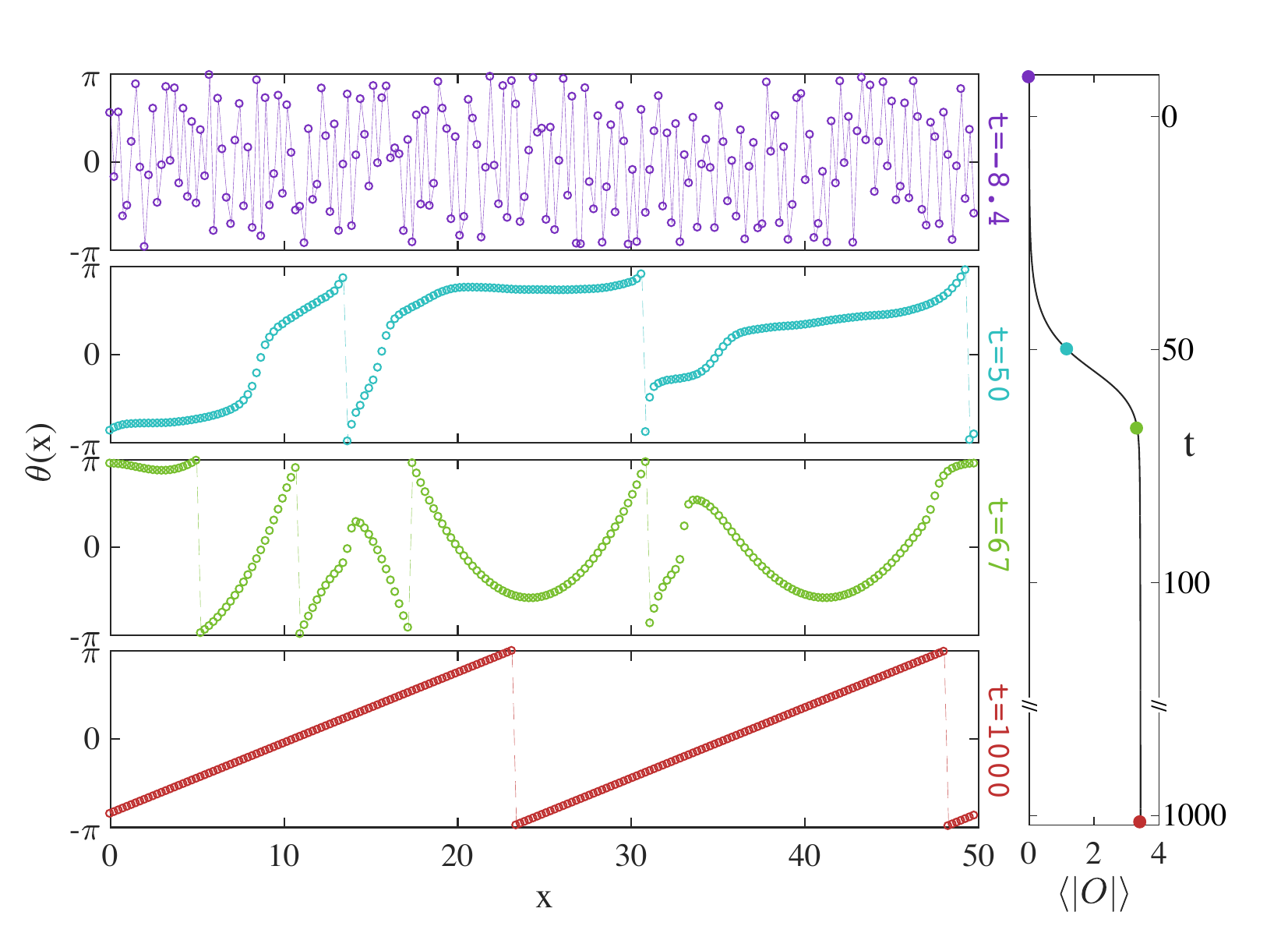}
\put(-238,150){(a)}
\hspace{0.5cm}
\includegraphics[trim=1.5cm 4.7cm 4.5cm 2.5cm, clip=true, angle=0, scale=0.25]{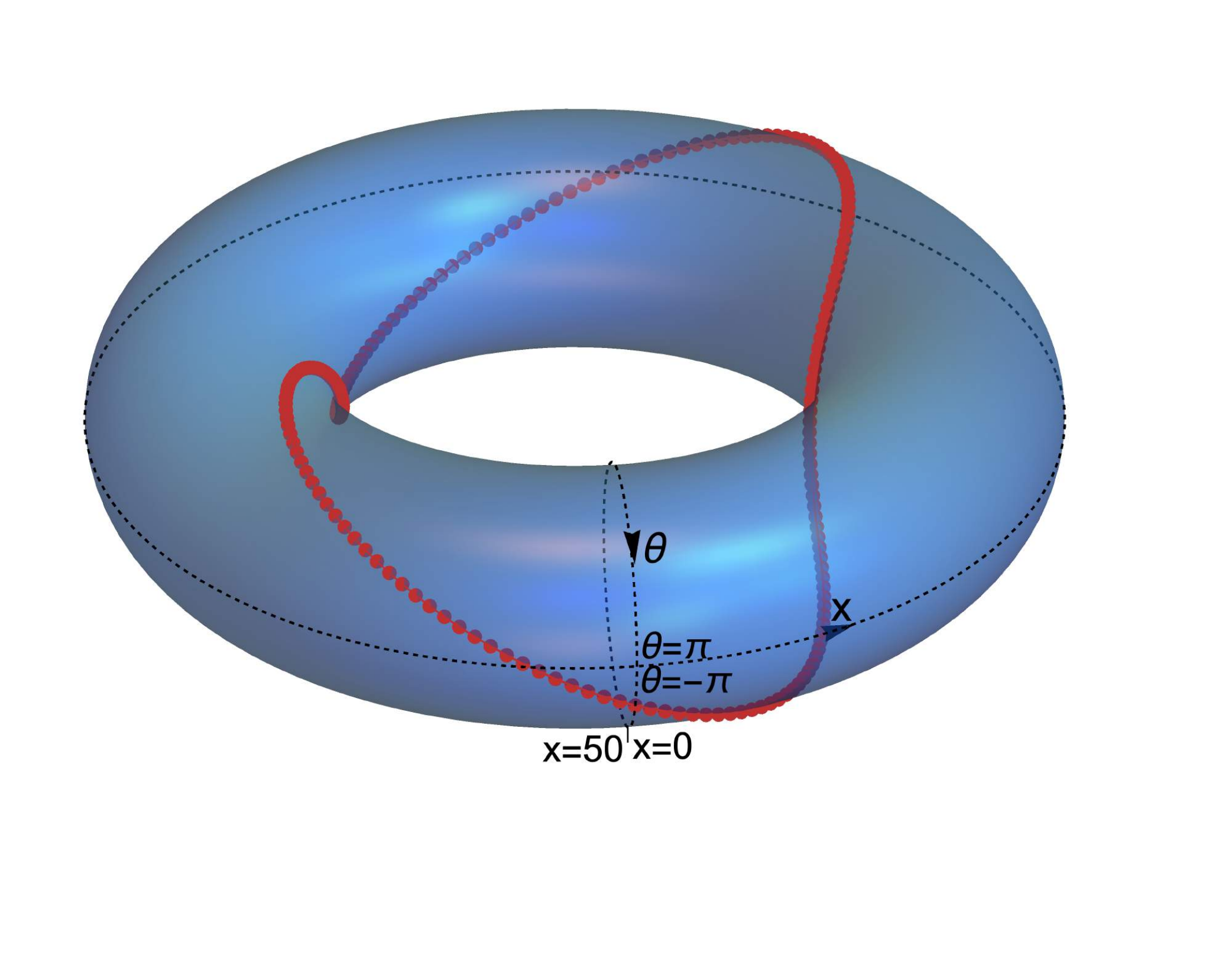}
\put(-220,150){(b)}
\caption{{\bf Time evolutions of the phases and the condensate of the order parameter in the quenched dynamics.} (a) Temporal evolutions of the phase $\theta$ from the initial time $(t=-8.4)$ to the final equilibrium state $(t=1000)$, and the corresponding evolution of the average condensate of the order parameter.  The phase evolves from initial random distributions to the final equilibrium state with constant gradient $\nabla \theta$ along the ring.  The dashed lines indicate spurious jumps of the phase at the edges $\theta=\pm\pi$. The winding number in the final equilibrium state is $W=+2$. The four specific instants are denoted in the right plot for the evolution of the order parameter condensate. (b) A stereographic view of the windings of the phase along the ring at time $t=1000$. The spurious jumps of the phase in panel (a) disappear on this $(\theta,x)$ torus, and the phase wraps the torus clock-wisely twice along the $x$-direction. In this figure, we have set $a_x=0$ and the quench rate $\tau_Q=20$. }\label{p1}
\end{figure}

 According to KZM, quenching the system across the critical point into the symmetry-breaking phase will result in the topological defects dynamically and statistically \cite{Kibble:1976sj,Kibble:1980mv,Zurek:1985qw}. In our model, the topological defect is the winding number of phase of the order parameter along the ring \cite{Sonner:2014tca, Das:2011cx}. (Details of the quench profile and numerical schemes can be referred to in the \hyperref[methods]{Methods}.)  We denote the compact ring as $\mathcal{C}$ with the circumference $L=50$, i.e., $x\in[0,50]$. Therefore, the winding number of the phase $\theta$ of the order parameter $\Psi_1=|\Psi_1|e^{i\theta}$ can be defined as 
\be\label{W}
W=\frac{1}{2\pi}\oint_\mathcal{C} {d{\bf \theta}}.
\ee
Fig.\ref{p1}(a) shows the temporal evolutions of the phases and average condensate of the order parameter from the initial time (at temperature $T=1.4T_c$) to the final equilibrium state ($T=0.8T_c$). At the initial time $t=-8.4$, the system is in the normal state with vanishing order parameter. Tiny random seeds of the scalar field are thrown into the system and evolves by keeping the temperature at $1.4T_c$ for a while in order to have a thermal equilibrium state at the initial time (refer to the right plot of Fig.\ref{p1}(a)).  Thus, the phase is randomly distributed in space at $t=-8.4$.  Decreasing the temperature across the critical point $T_c$, spontaneous U(1) symmetry breaking will take place and the winding numbers will turn out in the ring due to KZM. 

At the time $t=50$ the system is already in a superconducting state, but still in a far-from-equilibrium state which can be seen from the right plot in Fig.\ref{p1}(a). The phase $\theta$ at this stage has some roughly constant `plateaus', which is a direct result of KZM's prediction that the symmetry will spontaneously break and the phase will randomly choose some constant values in different spatial regions. Since the system is still in the far-from-equilibrium state, the winding numbers at this stage may be destroyed by the non-equilibrium dynamics for some reason. For instance, at $t=50$ the winding number is $W=+3$, however at the later time $t=67$ the winding number becomes $W=+2$.\footnote{We define the winding number to be $W=+n$ ($n\geq0$ and $n\in \mathbb{Z}$) if the phase goes from $-\pi$ to $+\pi$ and wraps it $n$ times along the $x$-direction. Negative winding number can be defined conversely. }  

The instant $t=67$ is at the early stage when the condensate of order parameter arrives at the equilibrium value. From $t=67$ to $t=1000$ the absolute value of the condensate of order parameter will keep invariant as well as the winding numbers. However, the phase of the order parameter still undergos dynamical processes until its gradient is a constant. For instance, at the final equilibrium state $(t=1000)$ the phase becomes `piecewise' straight lines which is different from that at $t=67$. This is from the fact that in the final equilibrium state, the superflow  has a constant velocity $\nabla\theta$ along the ring. The phase has a range $\theta\in[-\pi, \pi]$, therefore, the dashed lines in Fig.\ref{p1}(a) are just the spurious jumps of the phases at the edges $\theta=\pm\pi$. At the equilibrium state $\nabla\theta$ is a constant, thus from Eq.\eqref{W} it is easy to deduce that $\nabla\theta=2\pi W/L$. In fact the phase is smooth at these jumps if one views the phase in a 2-torus in the $(\theta,x)$-coordinates, refer to Fig.\ref{p1}(b). The phase smoothly wraps the torus clock-wisely twice if we go along the $x$-direction. Therefore, the phase has winding number $W=+2$ as we defined.

\begin{figure}[t]
\centering
\includegraphics[trim=1.5cm 0cm 1.5cm 1.cm, clip=true, scale=0.32]{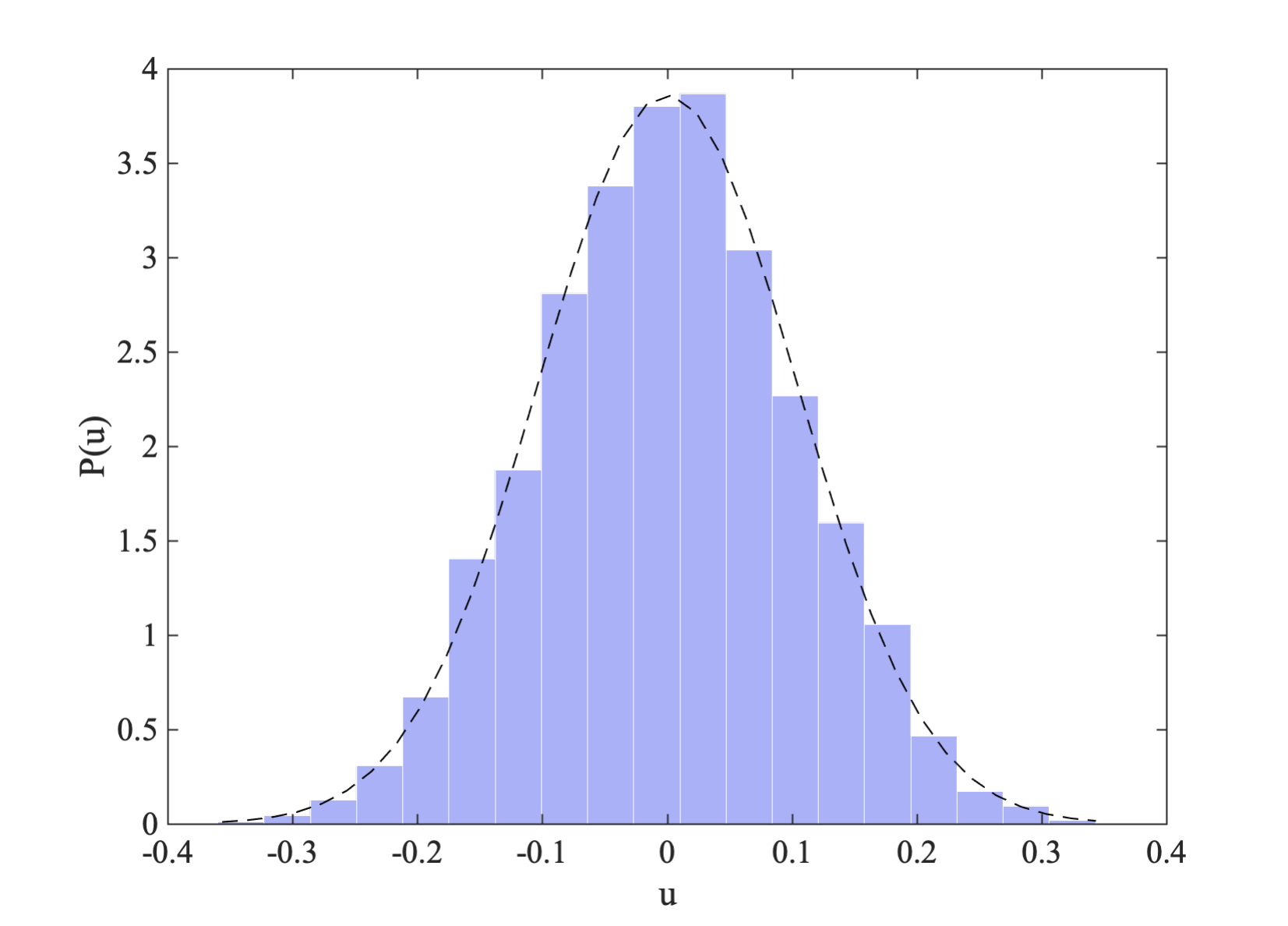}
\put(-233,168){(a)}~
\includegraphics[trim=0cm -0.3cm 1cm 0cm, clip=true, scale=0.3]{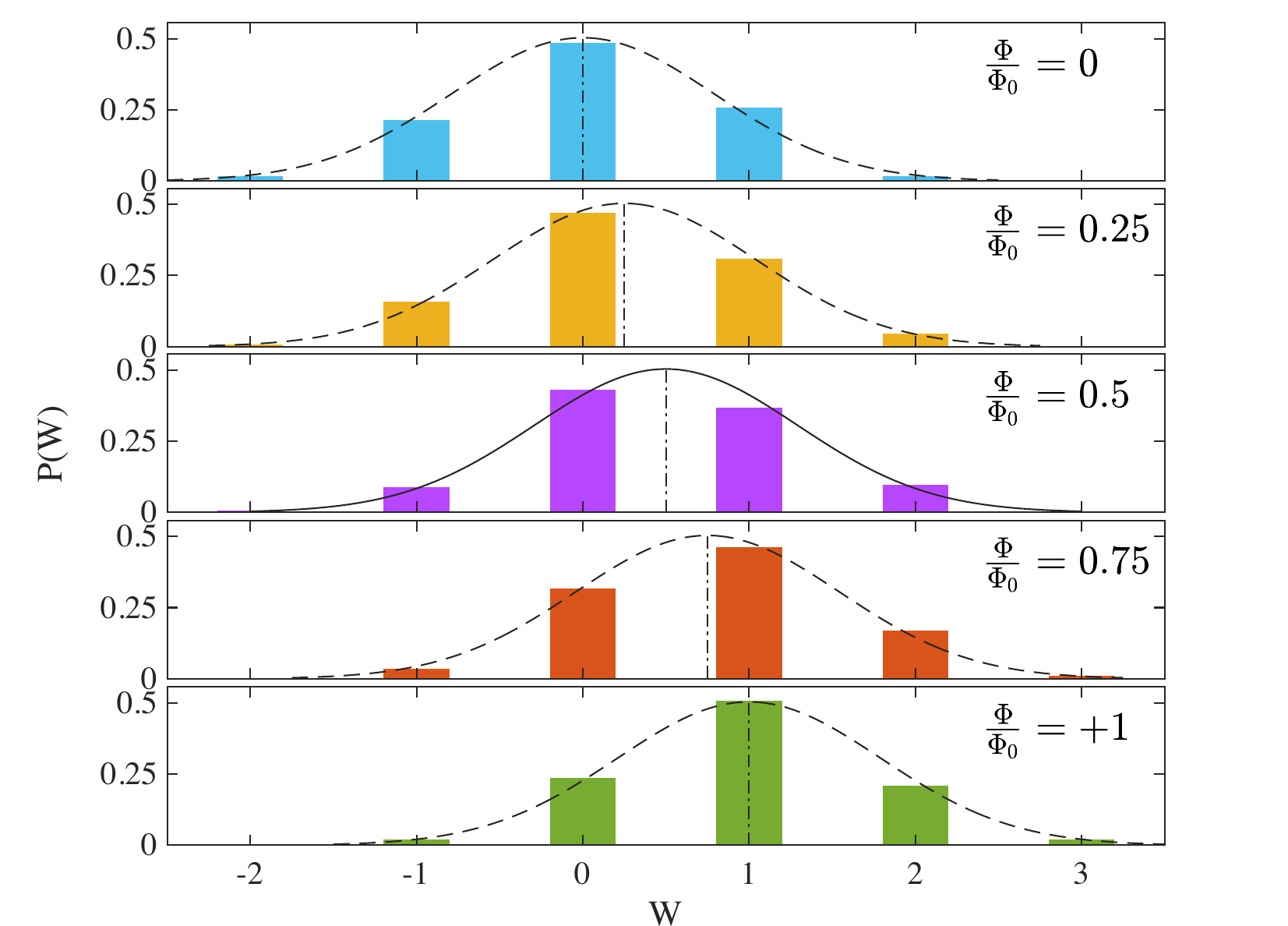}
\put(-230,168){(b)}
\caption{{\bf Distributions of the gauge invariant velocities and winding numbers.} (a) Probability densities of the gauge invariant velocity $\bf u$. The dashed line is the best fit to the numerical results, which satisfies the normal distribution with the mean $\langle u\rangle\approx 0$ and the standard deviation $\sigma({\bf u})\approx0.1033$. (b) Distributions of winding numbers (colored bars) for various magnetic fluxes. Dashed lines are the best fits to the numerical results. The dashed lines satisfy the normal distributions, with their means $\langle W\rangle\approx\Phi/\Phi_0$ as the dotted dashed lines indicate in each plot, and the standard deviations are identical to  $\sigma\approx0.7927$. We made $7000$ times of independent simulations in this figure.  }\label{p2}
\end{figure}

 \subsection{Statistics of winding numbers}
\label{dist} Near the boundary $z\to0$ we impose the Dirichlet boundary conditions for $A_x$ by fixing $a_x$. Integration of $a_x$ along $\mathcal{C}$ is the magnetic flux $\Phi$ threading the ring, 
\be\label{Phi}
\Phi=\int {\bf B}d{\bf S} =\int \left(\nabla\times{ a_x}\right)d{\bf S} =\oint_{\mathcal C}{a_x} d{x}=a_x L.
\ee
The last equality holds because we set a uniform $a_x$ along the ring. The gauge invariant velocity of the superflow is defined as ${\bf u}={\bf\nabla} \theta-a_x$ \cite{tinkham}.
For a fixed value of $a_x$ (equivalently to fix $\Phi/\Phi_0$ where $\Phi_0=2\pi$ is the flux quantum), we are able to get a distribution of winding numbers from numerous independent simulations according to the statistical properties of KZM \cite{Sonner:2014tca,delCampo:2021rak}. At the final equilibrium state, ${\bf u}=\nabla \theta-a_x=\frac{2\pi}{L}(W-\Phi/\Phi_0)$ from Eqs.\eqref{W} and \eqref{Phi}. Thus, as one varies $a_x$ and performs multiple simulations one can reach a distribution of ${\bf u}$.   Due to the `central limit theorem' \cite{Feller}, large number of independent simulations ($7000$ times in this paper) of $\bf u$ should satisfy the normal distribution, shown in Fig.\ref{p2}(a). The dashed line is the normal distribution function with the mean $\langle{\bf u}\rangle\approx 0$ and standard deviation $\sigma({\bf u})\approx 0.1033$. Numerical results (purple bars) match the normal distribution very well. We further check that the higher cumulants of ${\bf u}$, for instance, the third and fourth cumulants are respectively $\kappa_3({\bf u})\approx 4.9064\times10^{-6}$ and $\kappa_4({\bf u})\approx -1.6690\times10^{-5}$,  which are very tiny and numerically verifies that $P(\bf u)$ is a normal distribution \cite{cumulants}.   It should be noted that the winding numbers $W$ are integers, therefore, the distribution for $W$ can never be normal distribution. However, one can first assume $W$ as a continuous variable which satisfies the normal distribution as the dashed lines in Fig.\ref{p2}(b) shows. Then, the probabilities for each integer winding number are in fact the probabilities of the normal distribution at those integers. Just as Fig.\ref{p2}(b) shows, numerical results for $W$ (colored bars) match the dashed lines of the normal distributions very well at each integer.  The normal distributions in Fig.\ref{p2}(b) have the means $\langle W\rangle\approx\Phi/\Phi_0$ for each plot as the dotted dashed lines indicate. Their standard deviations are identical as $\sigma(W)\approx0.7927$. 

At the final equilibrium, $\langle W\rangle\approx\Phi/\Phi_0$ can be derived from $\langle {\bf u}\rangle\approx 0$ and ${\bf u}=\frac{2\pi}{L}(W-\Phi/\Phi_0)$. From the normal distribution, one can readily get the standard deviation of $W$ as $\sigma(W)=\sigma({\bf u})L/(2\pi)$. Theoretically, $\sigma(W)=0.1033\times50/(2\pi)\approx 0.8220$ which has $3.5\%$ difference to the numerical results $\sigma(W)\approx0.7927$. The small difference supports our conclusion above that the distribution of $W$ was constrained as probabilities at integers in the normal distribution. 


\begin{figure*}[t]
\centering
\includegraphics[trim=0cm 0.cm 1cm 0cm, clip=true, scale=0.3]{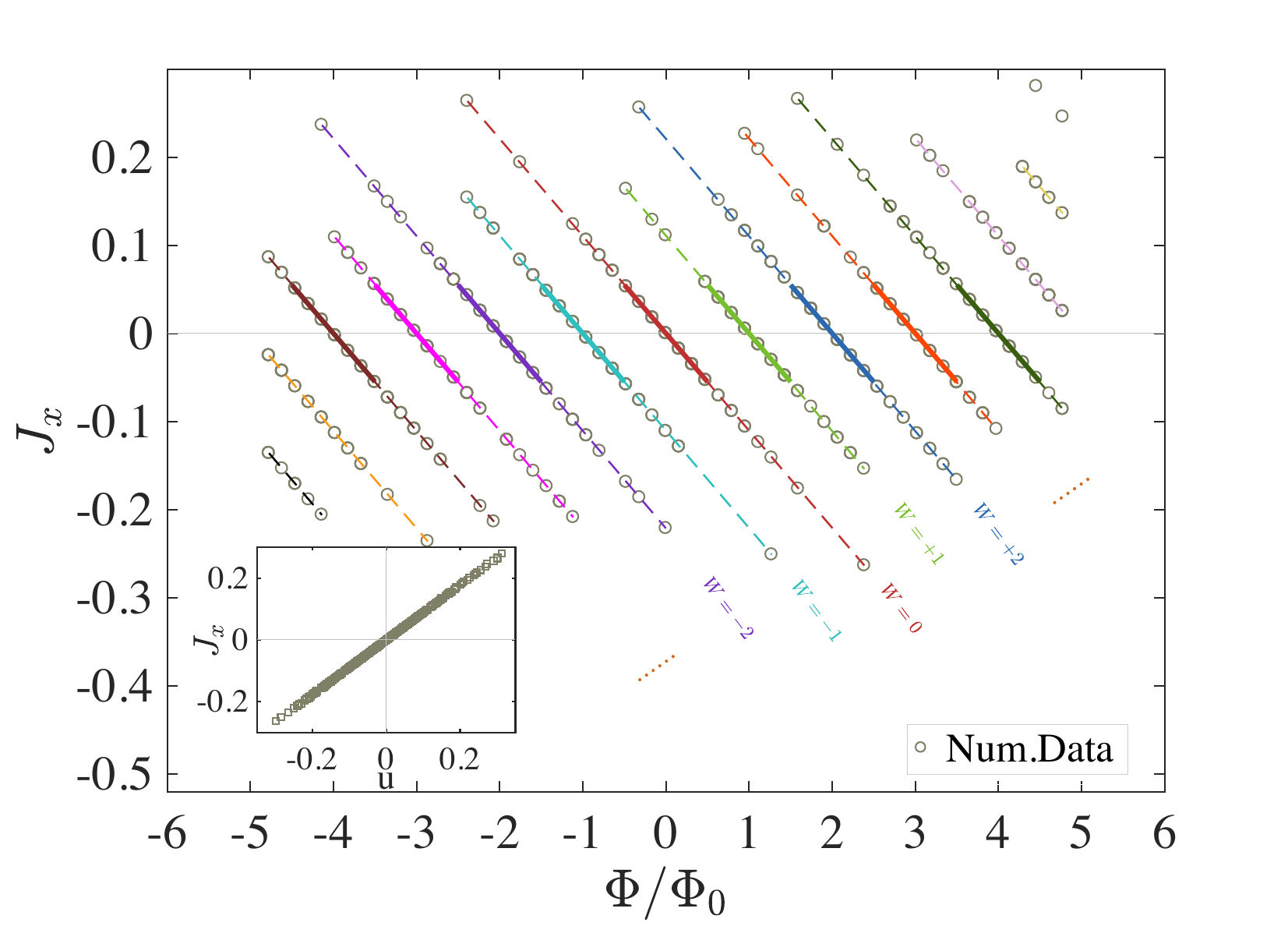}~~~
\includegraphics[trim=0.cm 0.cm 1cm 0cm, clip=true, scale=0.3]{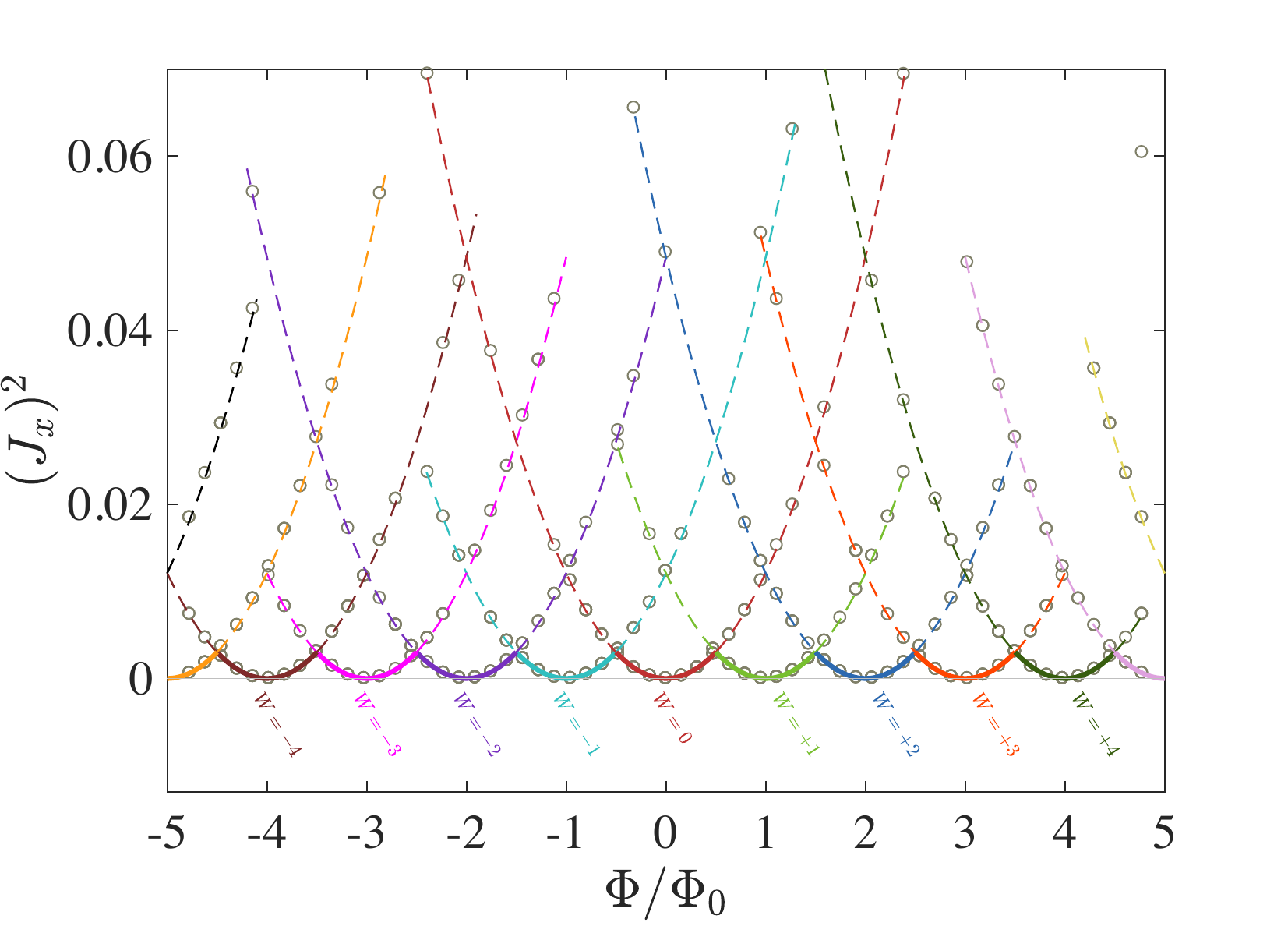}
\put(-465,155){(a)}
\put(-230,155){(b)}\\
\includegraphics[trim=0cm 0.cm 1cm 0cm, clip=true, scale=0.3]{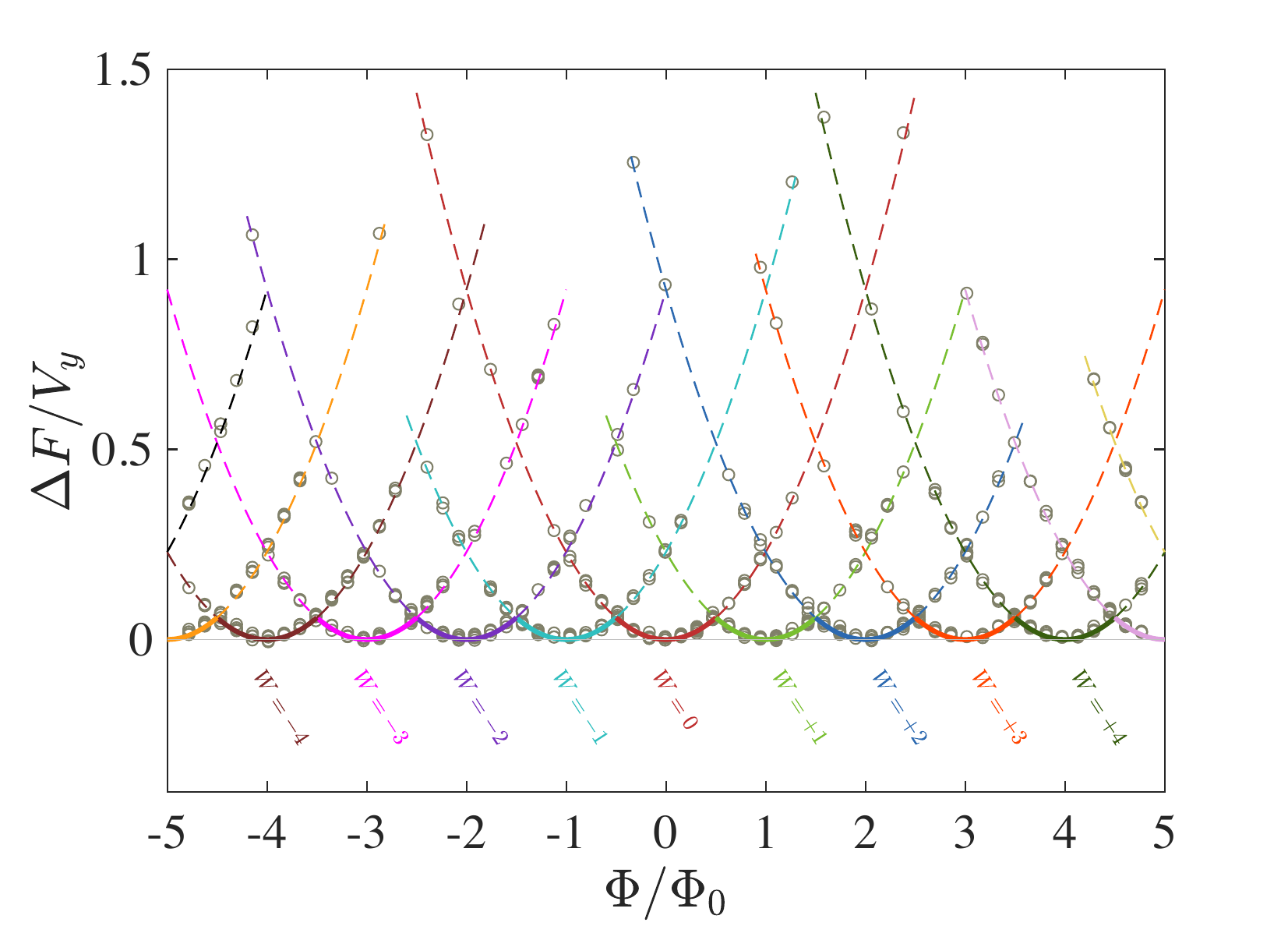}~~~
\includegraphics[trim=0.cm 0.cm 1cm 0cm, clip=true, scale=0.3]{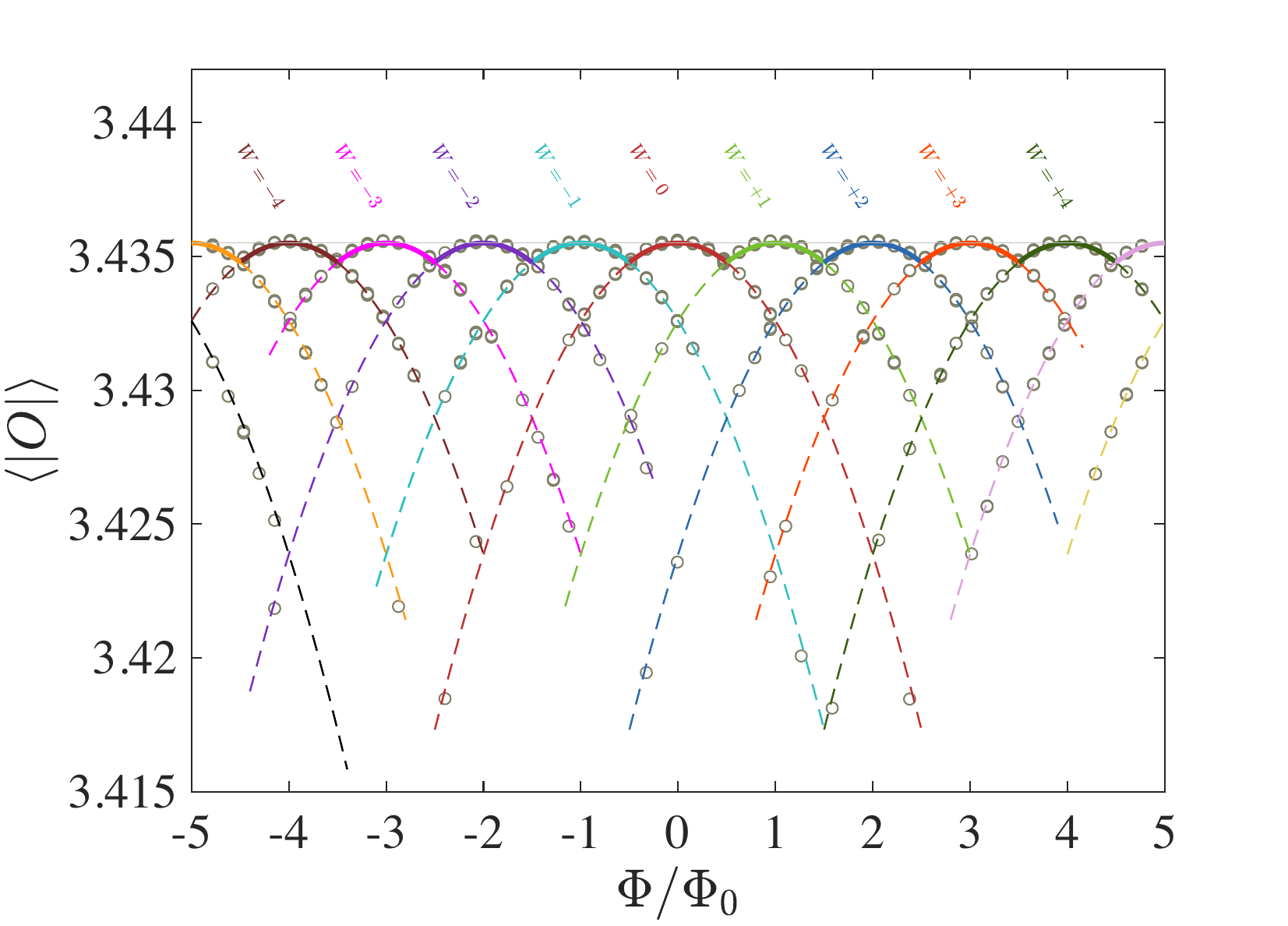}
\put(-465,155){(c)}
\put(-230,155){(d)}
\caption{{\bf Little-Parks periodicities with respect to the reduced magnetic flux $\Phi/\Phi_0$.} Panels (a),(b),(c) and (d) show the periodic dependences of the conserved current $J_x$, the square of the current $(J_x)^2$, the reduced free energies $\Delta F/V_y$ and the average condensate of the order parameter $\langle|O|\rangle$ versus the magnetic flux $\Phi/\Phi_0$, respectively. All the periods are $\Phi/\Phi_0=1$. Numerical results are indicated by the open circles. Each colored line is the best fit to the numerical results, and corresponds to specific winding number. The solid thick lines represent the favorable solutions corresponding to lower free energies. This also means dashed lines have higher free energies and are unfavorable. The inset plot in panel (a) shows a linear relation between the current $J_x$ and the gauge invariant velocity $\bf u$. }\label{p3}
\end{figure*}

\subsection{Little-Parks periodicities}
 From the AdS/CFT dictionary, the conserved current $J_x$ in the boundary field theory is $J_x=-b_x$.  The inset plot in Fig.\ref{p3}(a) shows a linear relation between $J_x$ and the gauge invariant velocity $J_x\propto\bf u$.  Since ${\bf u}=\frac{2\pi}{L}(W-\Phi/\Phi_0)$ at the equilibrium, we expect that $J_x$ has a periodic linear relation with respect to $\Phi/\Phi_0$ for each winding number, and the period is $\Phi/\Phi_0=1$. This relation is shown in Fig.\ref{p3}(a) with the open circles being the numerical data and the linear lines being the linear fitting $J_x\propto (W-\Phi/\Phi_0)$. Different colors stand for different winding numbers. The periodicity in Fig.\ref{p3}(a) is the result from the LP effect in a compact geometry \cite{tinkham}. The solid thick lines represent the parts which correspond to the minimal values of free energies (refer to Fig.\ref{p3}(c)).  Fig.\ref{p3}(b) shows the periodic relations between $(J_x)^2$ and magnetic flux $\Phi/\Phi_0$. The colored parabolas have the relation $(J_x)^2\propto(W-\Phi/\Phi_0)^2$. The solid thick lines have lower free energies than the dashed lines (see Fig.\ref{p3}(c)), therefore, a favorable solution between $(J_x)^2$ to the $\Phi/\Phi_0$ is reflected by the periodic solid lines, which is similar to the profiles in LP oscillations in the weakly coupled condensed matter physics \cite{tinkham}. 

Fig.\ref{p3}(c) exhibits the periodic parabolas of the reduced free energy $\Delta F/V_y$ with respect to $\Phi/\Phi_0$, in which $\Delta F=F-F_{W=0,\Phi/\Phi_0=0}$ and $V_y=\int dy$ is the volume along the $y$-direction. The solid bold lines stand for minimum values of free energies. The period is still $\Phi/\Phi_0=1$ and the lower free energies will change from one winding number to another winding number at half-integers of $\Phi/\Phi_0$. It is clear that the phase transition is of first order. For a specific value of $\Phi/\Phi_0$, a favorable solution is that with $W=[\Phi/\Phi_0]$, in which $[\Phi/\Phi_0]$ represents the integers closest to $\Phi/\Phi_0$. For instance, the closest integer to $\Phi/\Phi_0=0.75$ is $1$, thus the favorable solution has winding number $W=1$ which is indicated by the solid thick green line. It reminds us of the distributions of winding numbers in Fig.\ref{p2}(b) that for a fixed $\Phi/\Phi_0=0.75$ the most frequently appeared winding number is $W=1$ as well. In addition, if $\Phi/\Phi_0=0.5$, the closest integer to $0.5$ is either $0$ or $1$, thus the favorable solution has two choices with either $W=0$ or $W=1$. This is represented by the intersecting point between $W=0$ and $W=1$ solid lines. It was also reflected in Fig.\ref{p2}(b) that for a fixed $\Phi/\Phi_0=0.5$, the solutions $W=0$ and $W=1$ almost have the same probabilities. Thus, the free energy analysis is consistent with the statistical distribution of winding numbers in the preceding subsection, and we conjecture that the most frequently appeared solutions in statistics incline to have lower free energies in equilibrium. 

 Fig.\ref{p3}(d) shows the periodic parabolas of the average condensate of the order parameter versus $\Phi/\Phi_0$. It also has the period $\Phi/\Phi_0=1$ and the solid thick lines correspond to the lower free energy parts in Fig.\ref{p3}(c). The transit of the favorable solutions between different winding numbers take places at half-integers of $\Phi/\Phi_0$ as well. 
  
\section{Concluding remarks}
 Utilizing the KZM, we dynamically and statistically realized the winding numbers of the order parameter in a compact ring. At the final equilibrium state, the distribution of the gauge invariant velocity satisfies the normal distribution, while the distributions of integer winding numbers are constrained by a related normal distribution that the probability for a specific winding number equals the probability density in the normal distribution at that winding number.  LP periodicities turned out statistically as we varied the magnetic flux and simulated it multiple times. In particular, the conserved currents, the free energy and the average condensate of the order parameters exhibit the periodic dependences with respect to the magnetic flux.  They all have the periods as $\Phi/\Phi_0=1$. Favorable solutions with lower free energies display the scalloped shape of curves, and the scallops changes from one winding number to another winding number at the half-integers of $\Phi/\Phi_0$, implying a first order phase transition. Moreover, the analysis of the free energies were consistent with the statistics of winding numbers, which indicated that dynamically frequently appeared states incline to have lower free energies. Our work provided a possible avenue to study the lowest energy configurations of a quantum system from the dynamics and the statistics of the final states.

\section*{Acknowledgements}

 We appreciate the illuminating discussions with Han-Qing Shi. This work was partially supported by the National Natural Science Foundation of China (Grants No.11875095 and 12175008) and partially supported by the Academic Excellence Foundation of BUAA for PhD Students.

\appendix
\label{app}
\begin{center}
{\bf  Appendix: Little-Parks periodicities of static transition points from Sturm-Liouville eigenvalue problems }
\label{SL}\\
\end{center}

This part will analytically study the homogeneous and static holographic superconducting phase transition point in the main text by using the variational method for the Sturm-Liouville eigenvalue problems \cite{Siopsis:2010uq,Li:2011xja,Cai:2011ky}. It is convenient to work in the AdS-Schwarzschild black brane,
\begin{eqnarray}
ds^2=\frac{1}{z^2}\left(-f(z)dt^2+\frac{dz^2}{f(z)}+dx^2+dy^2\right) , ~~~~ \text{with}~ f(z)=1-\left(\frac{z}{z_h}\right)^3.
\end{eqnarray}

In the static case, the fields are homogeneous in the $x$-direction except that the phase of the order parameter will depend on $x$ along the compact ring \cite{Montull:2011im,Montull:2012fy,Cai:2012vk}.  Thus, we make ansatz as $\Psi=\Psi(z)e^{i\frac{2\pi W}{L}x}$, $A_t=A_t(z)$ and $A_x=A_x(z)$, where $W$ is the winding number and $L$ is the circumference of the ring. Therefore, the equations of motions read,
\begin{eqnarray}\label{eom1}
\left(\frac{1}{z^2}-z\right)\Psi''-\left(1+\frac{2}{z^3}\right)\Psi'-\left[\frac{m^2}{z^4}+\frac{A_t^2}{z^2(z^3-1)}+\frac{1}{z^2}\left(A_x-\frac{2\pi W}{L}\right)^2\right]\Psi&=&0,\\
\label{eom2}
A_t''+\frac{2\Psi^2e^{i\frac{4\pi W}{L}x}}{z^2(z^3-1)}A_t&=&0,\\
\label{eom3}
A_x''+\frac{3z^2}{z^3-1}A_x'+\frac{2\Psi^2e^{i\frac{4\pi W}{L}x}}{z^2(z^3-1)}A_x&=&0,
\end{eqnarray}
where $'$ denotes the derivative with respect to $z$. Near the boundary ($z \rightarrow 0$), we have the expansions,
\begin{eqnarray}
\Psi &=& \Psi_0 z^{\Delta_-} +\Psi_1 z^{\Delta_+} , ~~~ \text{where}~\Delta_\pm=\frac{3}{2} \pm \sqrt{\frac{9}{4}+m^2}\\
A_t &=& \mu-\rho z, \\
A_x &=& a_x+b_x z.
\end{eqnarray}
In the main text, we take $m^2=-2$ and focus on the standard quantization to set $\Psi_0=0$. Thus, the expectation values of the order parameter is $\langle O\rangle=\Psi_1$.  
At the transition point $T_c$ (or equivalently $\rho_c$), the scalar field is nearly vanishing $\Psi\sim0$. Therefore, the equation of motion Eq.\eqref{eom2} reduces to $A_t''\sim0$, with the solution $A_t\sim c_1+c_2 z$ where $c_{1,2}$ are constants. Imposing the vanishing boundary conditions of $A_t$ at the horizon $z=z_h=1$, one finally gets $A_t(z)=\rho(1-z)$. Similarly, one can get $A_x=a_x$ which is a constant from the Eq.\eqref{eom3} by imposing $\Psi\sim0$. 

Therefore, as $T\rightarrow T_c$ the equation Eq.(\ref{eom1}) for the scalar field $\Psi$ becomes
\begin{eqnarray}
-\Psi''+\frac{2+z^3}{z(1-z^3)}\Psi'+\left(\frac{-2+z^2\left(a_x-\frac{2\pi W}{L}\right)^2}{z^2(1-z^3)}-\frac{\rho^2}{(1+z+z^2)^2}\right)\Psi=0.
\end{eqnarray}
To solve this equation, we introduce a trial function $F(z)$ for $\Psi$ near $z=0$ as \cite{Siopsis:2010uq,Li:2011xja,Cai:2011ky}
\begin{eqnarray}
\Psi |_{z\rightarrow0} \approx \langle {O}\rangle z^2F(z)
\end{eqnarray}
The boundary condition for $F(z)$ is $F(0)=1$ and $F'(0)=0$.  Hence, the equation of motion for $F(z)$ is,
\begin{eqnarray}
-F''+\frac{1}{z}\left(\frac{2+z^3}{1-z^3}-4\right)F'+\left(\frac{4z^2+(a_x-\frac{2\pi W}{L})^2}{1-z^3}-\frac{\rho^2}{(1+z+z^2)^2}\right)F=0.
\end{eqnarray}
Multiplying
\begin{eqnarray}
K(z)=z^2(1-z^3)
\end{eqnarray}
to both sides of the above equation, the equation of motion for $F(z)$ reduces to
\begin{eqnarray}
\frac{d}{dz}\left(K(z)F'\right)+P(z)F+Q(z)\rho^2F=0
\end{eqnarray}
where 
\begin{eqnarray}
P(z)=4z^4+z^2\left(a_x-\frac{2\pi W}{L}\right)^2,~~~~
Q(z)=-\frac{z^2(1-z)}{1+z+z^2}
\end{eqnarray}
From the Sturm-Liouville eigenvlaue problem, the minimal eigenvalues of $\rho^2$ can be obtained by taking variations with the following functional \cite{Siopsis:2010uq,Li:2011xja,Cai:2011ky},
\begin{eqnarray}
\mu^2=\frac{\int_0^1dz(KF'^2+PF^2)}{\int_0^1dz QF^2}
\end{eqnarray}
The trial function $F(z)$ can be assumed to be $F(z)= 1-\alpha z^2$ with $\alpha$ a constant.
Thus, we obtain 
\begin{eqnarray}\label{mu2}
\rho^2(\alpha)=2\times\frac{1+\frac{1}{3}(a_x-\frac{2\pi W}{L})^2-\left[\frac{4}{3}+\frac{2}{5}(a_x-\frac{2\pi W}{L})^2\right]\alpha+\left[\frac{4}{5} +\frac{1}{7}(a_x-\frac{2\pi W}{L})^2 \right]\alpha^2}{3-\ln  3+\frac{\pi}{\sqrt{3}}+(\frac{13}{3}-4\ln 3)\alpha+(\frac{\pi}{\sqrt{3}}-\frac{7}{10}+\ln 3)\alpha^2 }.
\end{eqnarray}
\begin{figure}[h]
\centering
\includegraphics[trim=0cm 0cm 1cm 0.cm, clip=true, scale=0.4]{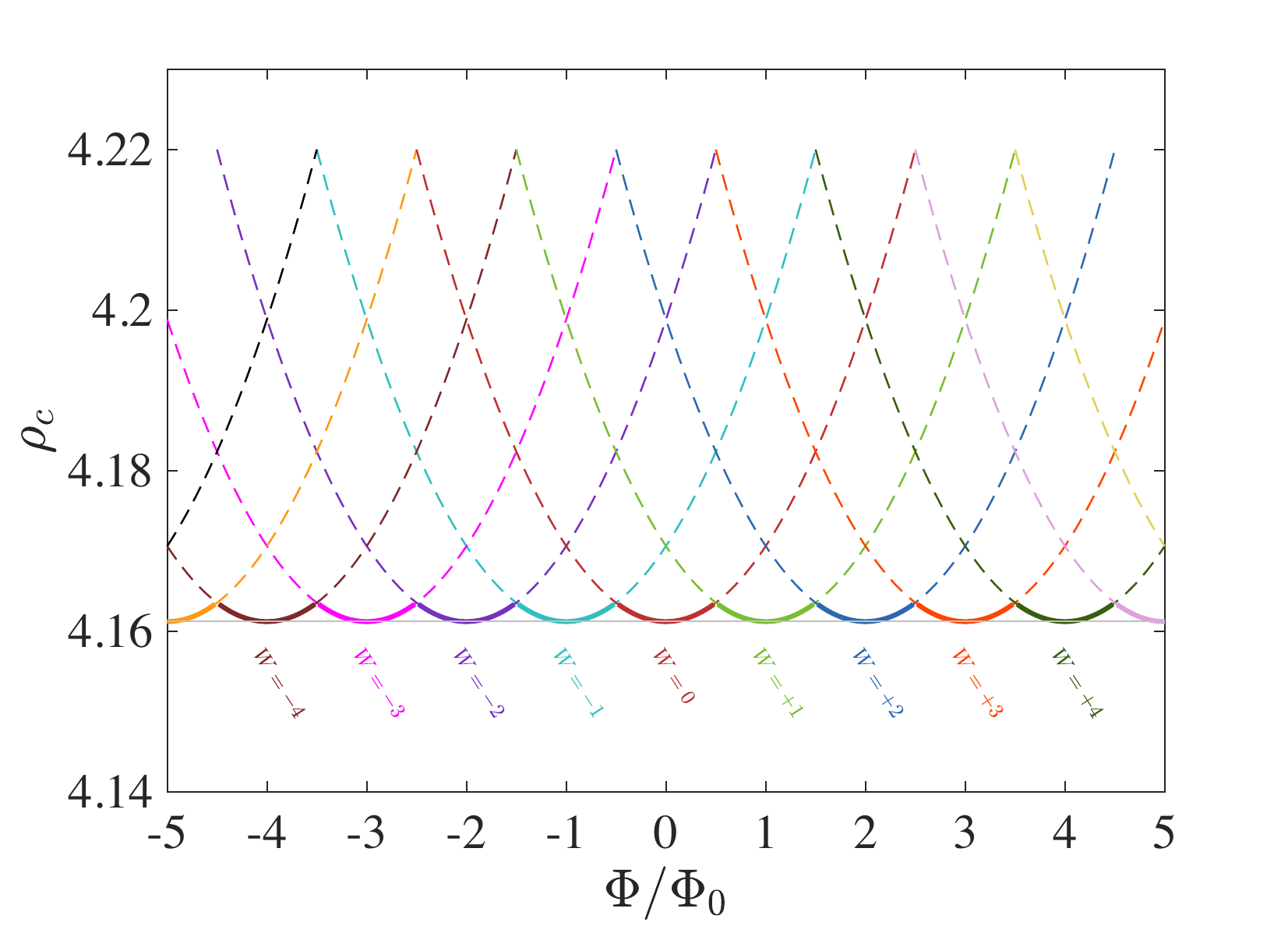}
\caption{Little-Parks periodic relations between $\rho_c$ and $\Phi/\Phi_0$. Each colored parabola corresponds to each winding number. The period is $\Phi/\Phi_0=1$. The solid thick curves represent lower free energy solutions, while the dashed lines correspond to unfavorable solutions. }\label{rhoc}
\end{figure}
In the main text, we set the circumference of the ring as $L=50$. In order to get the minimal value of $\rho$ (equivalently the critical point $\rho_c$), one needs to fix the values of $W$ and $a_x$ with $W\in\mathbb{Z}$. Thus one can find the critical value of $\rho_c$ from some specific values of $\alpha$ from the Eq.\eqref{mu2}. Then, one may sweep the values of $a_x$ and $W$ to find a series of values of $\rho_c$'s. This will provide a relation between the critical values of $\rho_c$ and $\left(a_x-\frac{2\pi W}{L}\right)$. As we mentioned in the main text, in the final equilibrium state (or at the static case), the magnetic flux $\Phi=a_x L$, therefore, $a_x-\frac{2\pi W}{L}=\frac{2\pi}{L}\left(\Phi/\Phi_0-W\right)$. The relations between the critical points $\rho_c$ and $\Phi/\Phi_0$ are given in Fig.\ref{rhoc}, with the period $\Phi/\Phi_0=1$.  Different colored lines correspond to different winding numbers, while the solid thick lines correspond to lower free energies in the main text.  The dashed lines correspond to unfavorable solutions. The scalloped shapes of the lower curves remind us of the Little-Parks effect \cite{LP,tinkham}. The transition of the lower curves from one winding number to another winding number happens at half-integers of $\Phi/\Phi_0$. 

We find that the quantity $\left(a_x-\frac{2\pi W}{L}\right)$ appear as a whole in Eq.\eqref{mu2}. Therefore, it is easy to deduce that the relation between the critical points $\rho_c$ should be a function of $\left(a_x-\frac{2\pi W}{L}\right)$.   By fitting the data in Fig.\ref{rhoc} we find that it is a parabolic relation as
\begin{eqnarray}
\rho_c\approx a_1 \left(\frac{L}{2\pi}a_x-W\right)^2+\rho_{\text{min}}=a_1\left(\Phi/\Phi_0-W\right)^2+\rho_{\text{min}}, 
\end{eqnarray}
with $a_1\approx0.0094$ and $\rho_{\text{min}}\approx4.16$.


\end{document}